\documentclass[letter]{aa} 

\usepackage{graphicx}
\usepackage{threeparttable}
\usepackage{wasysym}

\usepackage{txfonts}
\begin{document}

   \title{Assessing the residual biases in high-resolution transit absorption spectra correction}

   \author{W. Dethier\inst{\ref{1}} \and B. Tessore\inst{\ref{2}}
          }
   \institute{Instituto de Astrofisica e Ciencias do Espaco, Universidade do Porto, CAUP, Rua das Estrelas, 4150-762 Porto, Portugal\label{1} \\ 
              \email{william.dethier@astro.up.pt} \\
              \and Rue Menon, 38000, Grenoble, France\label{2}            
              }

   \date{}

  \abstract
   {In recent years it has become common practice to divide observed transit absorption spectra by synthetic absorption spectra computed for the transit of an atmosphere-less planet. This action supposedly corrects the observed absorption spectrum, leaving the sole atmospheric absorption signature free from the biases induced by stellar rotation and centre-to-limb variations.}
   {We aim to show that while this practice is beneficial, it does not completely correct the absorption spectrum from the stellar distortions and that some residual biases remain, leaving a possibly altered atmospheric signature. }
   {By reducing the problem to its most basic form, we show that dividing the observed absorption spectrum by a synthetic absorption spectrum of the planet does not isolate the pure atmospheric absorption signature. We also used simulated synthetic transit observations to assess the magnitude of these residual biases in typical transit observations. }
   {We show that dividing the observed absorption spectrum by the planetary absorption spectrum results in an atmospheric signature modulated by the ratio of the flux behind the atmosphere and the flux behind the planet. Depending on the non-homogeneity of the stellar spectrum, this leads to distorted atmospheric signatures. Eventually, directly analysing these biased signatures will lead to wrong estimates of planetary atmosphere properties.}
   {}

   \keywords{Methods: data analysis -- Techniques: spectroscopic -- Planets and satellites: atmospheres
               }

   \maketitle
%

\section{Introduction}
It is now commonly known that high-resolution transit absorption spectra are contaminated by planet-occulted line distortions (POLDs) originating from the non-homogeneity of the stellar spectrum over the surface of a star. These POLDs can bias the absorption spectra to a point where the distortions can be interpreted as, or even hide, actual absorption signatures of a planetary atmosphere.
A way to correct these biases regularly seen in the literature in recent years is to divide the observed absorption spectra by a synthetic absorption spectrum computed from the transit of an opaque planetary disc \citep[e.g.][]{Casasayasbarris2017, casasayasbarris2018, yan2018,nugroho2020,maguire2023,keles2024,maguire2024,masson2024}. In this way, only the POLDs are included in the simulated absorption spectra, and the division is thought to leave only the absorption signature of the atmosphere. This residual signal can then be fitted and used to derive atmospheric properties. However, with such a correction, the absorption spectra are not fully corrected from the POLDs and residual biases remain, affecting the true atmospheric signal. In this Letter we aim to show how absorption spectra remain biased by the stellar spectrum even after correction, in addition to assessing the magnitude of these residual biases.
\section{Correcting absorption spectra}
\label{sec:2}
\begin{figure}
    \centering
    \includegraphics[width = 1\linewidth]{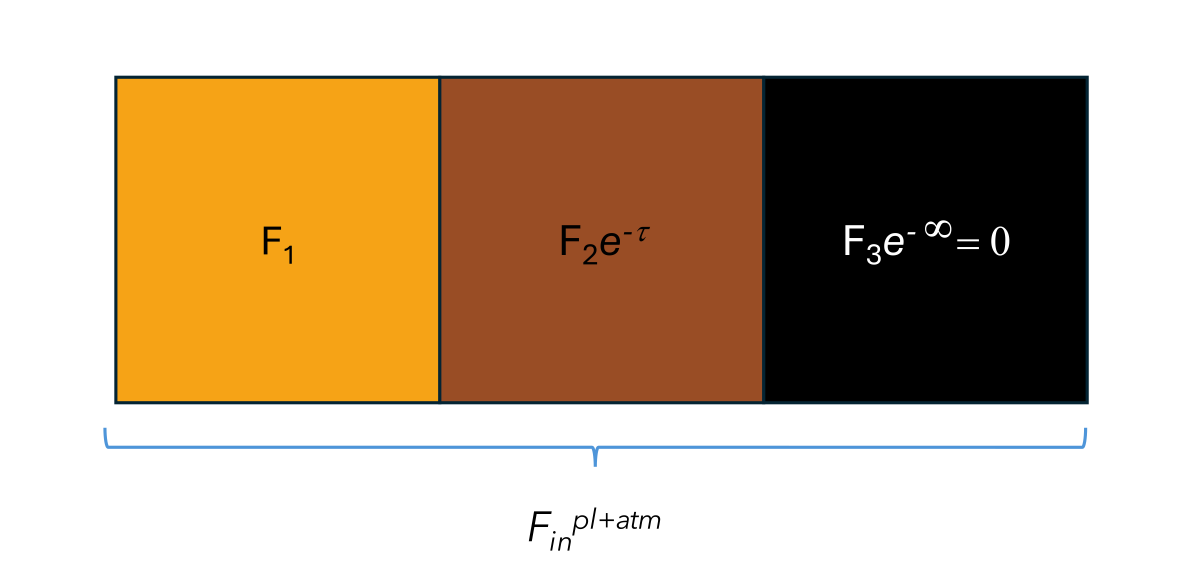}
    \caption{Three-cell star occulted by a planet and its atmosphere.}
    \label{fig:1}
\end{figure}

Let us assume a star made of three squares with respective fluxes $F_1$, $F_2$, and $F_3$ (see Fig. \ref{fig:1}) and a planet with an atmosphere transiting the star. Square 1 is unocculted, square 2 is completely occulted by the planetary atmosphere, and square 3 is completely occulted by the planet's opaque body (see Fig. \ref{fig:1}).
We then have the out-of-transit spectrum given by
\begin{equation}
\rm F_{out}(\lambda) = F_1(\lambda) + F_2(\lambda) + F_3(\lambda)
\end{equation}
and the in-transit spectrum given by
\begin{equation}
\rm F_{in}^{pl+atm}(\lambda) = F_1(\lambda) + F_2(\lambda) \times e^{-\tau(\lambda)}
,\end{equation}
where $\rm F_i$ is the local flux coming from the stellar cells, $\lambda$ is the wavelength dependence, and $e{(-\tau)}$ shows that during the transit $F_2$ has been absorbed by the planetary atmosphere, with $\tau$ the optical depth of the atmosphere.
The absorption spectrum is given by
\begin{equation}
\rm \mathcal{A}_{pl+atm}(\lambda)=\frac{F_{out}(\lambda) - F_{in}^{pl+atm}(\lambda) }{F_{out}(\lambda)}= \frac{F_2(\lambda)\times(1-e^{-\tau(\lambda)}) + F_3(\lambda)}{F_1(\lambda) + F_2(\lambda) +F_3(\lambda)}
\label{eq:3}
.\end{equation}
In the following equations we have dropped the wavelength dependence subscript for clarity.\\
\indent In the literature, it is common to simulate the absorption spectrum using only the opaque planetary body. The observed absorption spectrum is then divided by this simulated absorption spectrum to correct for the POLDs and supposedly derive the absorption spectrum of the planetary atmosphere. The residual is often fitted, and atmospheric properties are then derived from this signal.
The simulated absorption spectrum of the opaque planetary body, in our example, can be written as
\begin{equation}
\rm \mathcal{A}_{pl}=\frac{F_{out} - F_{in}^{pl} }{F_{out}}= \frac{F_3}{F_1 + F_2 +F_3}
\label{eq:4}
.\end{equation}
Dividing Eq. \ref{eq:3} by Eq. \ref{eq:4}, we get 
\begin{equation}
\begin{split}
\rm \dfrac{\mathcal{A}_{pl+atm}}{\mathcal{A}_{pl}}&=\frac{F_2\times(1-e^{-\tau}) + F_3}{F_3} \\
&= 1 + (1-e^{-\tau})\dfrac{F_2}{F_3}
\end{split}
\label{eq:5}
.\end{equation}
After this supposed correction, a factor $F_2/F_3$ still affects the atmospheric signature, $e^{-\tau}$. 

\subsection{Generalisation to a discretised stellar disc}
\label{sec:3}
In this section we generalise the derivation described above to a number of stellar cells, $n$. The out-of-transit spectrum is given by
\begin{equation}
\begin{split}
\rm \text{F}_\text{out} = &  \displaystyle\sum_{i=1}^n \text{F}_i
\label{eq:fout} 
\end{split}
,\end{equation}
where the $i$ index runs over all the cells of the stellar disc. The in-transit spectrum is given by
\begin{equation}
 \text{F}_{\text{in}}^{\text{pl+atm}}= \displaystyle\sum\limits_{i=1}^n\text{F}_i e^{-\tau_i}   \ ,
\label{eq:fin}
\end{equation}
with $\tau_{i}$ the spectral optical depth of the atmospheric column occulting the stellar cell, $i$. It is equal to 0 ($e^{-\tau_{i}}=1$) when there is no atmosphere or planetary opaque body in front of the stellar cell, $i$. It tends to infinity ($e^{-\tau_{i}}=0$) for all wavelengths when the cell is behind the planetary disc. It also tends to infinity for specific wavelengths when the cell is behind a part of the atmosphere that is optically thick.

The absorption spectrum from the planet and the atmosphere is given as
\begin{equation}
\begin{split}
& \rm \mathcal{A}_{pl+atm} = \frac{\overbrace{\displaystyle\sum\limits_k^K \text{F}_k}^\text{{\large$\text{F}_\text{abs pl}$}}   \ +\overbrace{\displaystyle\sum\limits_l^L \text{F}_l\ (1-e^{-\tau_l})}^\text{{\large $\text{F}_\text{abs atm}$}}}{\displaystyle\sum_{i=1}^n \text{F}_i},
\end{split}
\label{eq:fout_fin}
\end{equation}
where the $k$ index runs over the stellar cells that are occulted by the planetary opaque disc at a specific timestep, and the $l$ index runs over the stellar cells occulted by the planetary atmosphere at a specific timestep.
The absorption spectrum by the opaque planetary body is given by
\begin{equation}
\begin{split}
& \rm \mathcal{A}_{pl} = \frac{\overbrace{\displaystyle\sum\limits_k^K \text{F}_k}^\text{{\large$\text{F}_\text{abs pl}$}}}{\displaystyle\sum_{i=1}^n \text{F}_i}.
\end{split}
\label{eq:fout_fin1}
\end{equation}
Dividing Eq. \ref{eq:fout_fin} by Eq. \ref{eq:fout_fin1} to isolate the absorption spectrum of the atmosphere, we get
\begin{equation}
\begin{split}
 \rm \dfrac{\mathcal{A}_{pl+atm}}{\mathcal{A}_{pl}} =& \frac{\displaystyle\sum\limits_k^K \text{F}_k +\displaystyle\sum\limits_l^L \text{F}_l\ (1-e^{-\tau_l})}{\displaystyle\sum\limits_k^K \text{F}_k}\\
& =1 + \frac{\displaystyle\sum\limits_l^L \text{F}_l\ (1-e^{-\tau_l})}{\displaystyle\sum\limits_k^K \text{F}_k}.
\end{split}
\label{eq:fout_fin2}
\end{equation}
Under the assumption of an atmosphere with a constant optical depth surrounding the planet, the equation becomes
\begin{equation}
\begin{split}
 \rm \dfrac{\mathcal{A}_{pl+atm}}{\mathcal{A}_{pl}}& =1 + (1-e^{-\tau})\frac{\displaystyle\sum\limits_l^L \text{F}_l}{\displaystyle\sum\limits_k^K \text{F}_k}.
\end{split}
\label{eq:fout_fin3}
\end{equation}
This means that the true atmospheric absorption signature is modulated by the ratio of the flux occulted by the atmosphere and the flux occulted by the planet. With centre-to-limb variation, stellar rotation, and other phenomena, inducing local variation in the stellar spectrum over the surface of the star, the flux behind the planet is expected to differ from the flux behind the atmosphere, especially for broad atmospheres and large planets. Therefore, one can expect the flux ratio to deform the true atmospheric absorption signature. 

If we consider the flux to be constant over the whole stellar surface, then Eq. \ref{eq:fout_fin3} becomes
\begin{equation}
\begin{split}
 \rm \dfrac{\mathcal{A}_{pl+atm}}{\mathcal{A}_{pl}}& =1 + (1-e^{-\tau})\dfrac{\rm S_{atm}}{\rm S_{pl}} ,
\end{split}
\label{eq:fout_fin4}
\end{equation}
with $\rm S_{pl}$ and $\rm S_{atm}$ being the projected surface of the opaque apparent planetary disc and the projected surface of the atmospheric annulus, respectively.\\
\indent However, in the case with constant flux over the stellar surface and constant planetary atmospheric optical depth, the absorption spectrum, Eq. \ref{eq:fout_fin}\footnote{This simplification comes from the definition of the flux coming from a single stellar cell: $\rm F_{cell} = I_{cell} \times \Delta\Omega_{cell} = I_{cell}\times S_{cell} / R_*^2 $, with $\rm I_{cell}$ the intensity coming from the stellar cell and $\rm \Delta\Omega_{cell}$ the solid angle subtended by the stellar cell.}, becomes\begin{equation}
\begin{split}
& \rm \mathcal{A}_{pl+atm} = \frac{\rm S_{pl}   \ + S_{atm}\ (1-e^{-\tau})}{\rm S_*},
\end{split}
\label{eq:fout_fin_simplified}
\end{equation}
where $\rm S_*$ is the projected surface of the apparent stellar disc.
To make it equivalent to Eq. \ref{eq:fout_fin4} and keep the same line-to-continuum ratio, one can multiply Eq. \ref{eq:fout_fin_simplified} by the surface ratio of the star to the planet: 
\begin{equation}
\begin{split}
& \rm \mathcal{A}_{pl+atm} \times  \frac{\rm S_{*}}{\rm S_{pl}} = 1+\ (1-e^{-\tau})\frac{S_{atm}}{\rm S_{pl}} = \dfrac{\mathcal{A}_{pl+atm}}{\mathcal{A}_{pl}}.
\end{split}
\label{eq:fout_fin_simplified_b}
\end{equation}

A more formal expression for the bias, $\mathcal{B_A}$, can be given by the difference between Eq. \ref{eq:fout_fin2} and Eq. \ref{eq:fout_fin_simplified_b},
\begin{equation}
\begin{split}
\rm \mathcal{B_A} \equiv & \rm \dfrac{\mathcal{A}_{pl+atm}}{\mathcal{A}_{pl}} - \left.\dfrac{\mathcal{A}_{pl+atm}}{\mathcal{A}_{pl}}\right|_{sim} \\
& = \frac{\displaystyle\sum\limits_l^L \text{F}_l\ (1-e^{-\tau_l})}{\displaystyle\sum\limits_k^K \text{F}_k} - \sum\limits_l^L(1-e^{-\tau_l})\frac{\rm S_l}{\rm S_{pl}}\\
& = \sum\limits_l^L(1-e^{-\tau_l}) \left[ \rm \frac{F_l}{F_{pl}} - \frac{S_l}{S_{pl}} \right]
\end{split}
\label{eq:bias}
,\end{equation}
where the optical depth in Eq. \ref{eq:fout_fin_simplified_b} was included back in the sum, $\rm \displaystyle\sum\limits_k^K F_k = F_{pl}$ and $ \rm \displaystyle\sum\limits_l^L S_l = S_{atm}$. The `sim' subscript refers to the simplified form of the equation (Eq. \ref{eq:fout_fin4}) when considering uniform stellar spectra and a constant atmospheric optical depth. If we use the setup from Fig. \ref{fig:1}, we can write the bias as
\begin{equation}
\begin{split}
\rm \mathcal{B_A} &= (1-e^{-\tau}) \left[ \rm \frac{F_{atm}}{F_{pl}} - \frac{S_{atm}}{S_{pl}} \right] = (1-e^{-\tau}) \left[ \rm \frac{F_{atm}}{F_{pl}} - \frac{R_{atm}^2 - R_{pl}^2}{R_{pl}^2} \right]  \\
&= (1-e^{-\tau}) \left[ \rm \frac{F_{atm}}{F_{pl}} - (\alpha^2 -1 ) \right]
\end{split}
\label{eq:bias2}
\end{equation}
if we assume $ \rm R_{atm} = \alpha R_{pl}$, with $\alpha > 1$, $\rm F_2 = F_{atm}$, and $ \rm F_3=F_{pl}$.
\section{Estimates of the residual biases: A simplified framework}
In this section we use transit simulations of a simplified planetary system setup to illustrate the possible biases to be expected on the derived planetary atmosphere parameters when correcting transit absorption spectra via division. 
\subsection{The star}
We simulated the star as a disc discretised by a two-dimensional square grid that we filled with synthetic spectra, accounting for centre-to-limb variations of the lines' profiles, broadband limb darkening, and stellar rotation. To simulate the synthetic spectra, we used the non-local thermodynamic equilibrium version of the Turbospectrum code for spectral synthesis \citep{plez1998,plez2012, heiter2021, larsen2022, magg2022}, along with a MARCS photospheric model \citep{gustafsson2008} for the temperature, log $g$, and metallicity of the star HD 209458 (see Table \ref{tab:tab1}), the host star of a well-known exoplanet \citep{henry2000,charbonneau2000}, and line lists from the VALD3 database \citep{Ryabchikova2015}. We restricted ourselves to a wavelength region around the sodium doublet in the visible to study the impact on a typical atmospheric sodium signature in a transiting close-in exoplanet. 
\begin{table}
\renewcommand{\arraystretch}{1.}
\setlength{\tabcolsep}{22pt}
\caption{Stellar and planetary properties. }
\label{tab:tab1}
\centering
\begin{threeparttable}
\begin{tabular}{l c}
\hline\hline
Parameter &  Value\\ 
\hline
\\
Stellar radius (R$_{\astrosun}$) & 1.155  \\
T$_{\text{eff}}$ (K) & 6\,065  \\
Metallicity [Fe/H] & 0.00 \\
log \textit{g} (cm s$^{-2}$) & 4.361 \\
Planetary radius (R$_\text{*}$) & 0.1145  \\
\\
\hline                                  
\end{tabular}
\begin{tablenotes}
\item[] \tiny{\textbf{Note.} The values are the same as those used or derived in \citet{casasayasbarris2020}. } 
\end{tablenotes}
\end{threeparttable}
\end{table}
\subsection{The transiting planet and atmosphere}
We studied 12 different cases for which we varied the position of the planet, the size of the atmosphere, and the $\rm v_{eq}\sin(i)$ of the star\footnote{$\rm v_{eq}$ is the rotational velocity of the star at its equator, and $i$ is the inclination of the stellar rotation axis.}. Table \ref{tab:2} shows the values that we used in the simulations for a specific atmosphere size. We used two sizes: an atmospheric radius of 1.5 times or 2 times the planetary radius of HD 209458 b, which is about 0.1145 times the radius of HD 209458. Once the spectra were distributed on the stellar grid, the out-of-transit spectrum was computed by summing the spectra from all the cells. To generate the in-transit spectra, we set the spectra from stellar cells behind the planet to zero and multiplied the spectra from the stellar cells behind the planetary atmosphere by $e^{-\tau}$ before computing the sum. For simplicity, we considered a homogeneous optical depth, $\tau$, for the whole atmosphere. The optical depth\footnote{Note that in the equations of Sect. \ref{sec:2}, we drop the index for the wavelength dependence of the optical depth.} is defined as
\begin{align}
\tau(\lambda) &=   \frac{\pi e^2}{m_e c}  \, f_{\text{osc}} \,\phi(\lambda - \lambda_0, w)\, n_\text{col,}
\end{align}
with $\phi(\lambda - \lambda_0, w)$ the line profile (we used a Voigt profile in this study), $n_\text{col}$ the column density, $w = \displaystyle \sqrt{\frac{2k_b T }{ m } }$ the thermal broadening, $T$ the temperature of sodium atoms, $f_{\text{osc}}$ the oscillator strength of the transition, $k_b$ the Boltzmann constant, $m$ the mass of a sodium atom, $e$ the electron electric charge, $m_e$ its mass, and $c$ the speed of light; $\lambda_0$ is the wavelength of reference for the atomic transitions of interest, here the Na I D2 line. No velocity field was included in the atmosphere. In the following simulations the temperature of the atmosphere was set to 10\,000K and the column density was set to 7$\times 10^{11}$ at. cm$^{-2}$. This allowed us to produce a sodium atmospheric absorption signature of about 0.6 and 1.4 $\%$ for the case with R$_{\rm atm} = 1.5 \ \rm R_{pl}$ and R$_{\rm atm} = 2.0 \ \rm R_{pl}$, respectively (see the black curves in Figs. \ref{fig:3} and \ref{fig:4}).  

\begin{table}
\renewcommand{\arraystretch}{1.5}
\setlength{\tabcolsep}{12pt}
\caption{List of the simulations.}
\label{tab:2}
\centering
\begin{threeparttable}
\begin{tabular}{ccc}
\hline
\hline
Case & $ \mathrm{v_{eq}\sin(i_*)\ [km\,s^{-1}]}$ & $x_\mathrm{pl}; y_\mathrm{pl}$  \\
\hline
1 & 4 & 0.0 ; 0.0 \\
2 & 4 & -0.5 ; 0.0 \\
3 & 10 & 0.0 ; 0.0\\
4 & 10 & -0.5 ; 0.0 \\
5 & 20 & 0.0 ; 0.0 \\
6 & 20 & -0.5 ; 0.0 \\
\hline
\end{tabular}
\begin{tablenotes}
\item[] \tiny{\textbf{Note.} The planet's positions are given in Cartesian coordinates in units of the stellar radius, with $(x_\mathrm{pl}; y_\mathrm{pl}) = (0.0;0.0)$ being the centre of the apparent stellar disc. The Cartesian $x$ and $y$ axes are aligned with the stellar equator and rotation axis, respectively.}
\end{tablenotes}
\end{threeparttable}
\end{table}
\subsection{Results of the simulations}
\begin{figure}
\hspace*{-5mm}
    \centering
    \includegraphics[width = 1.1\linewidth]{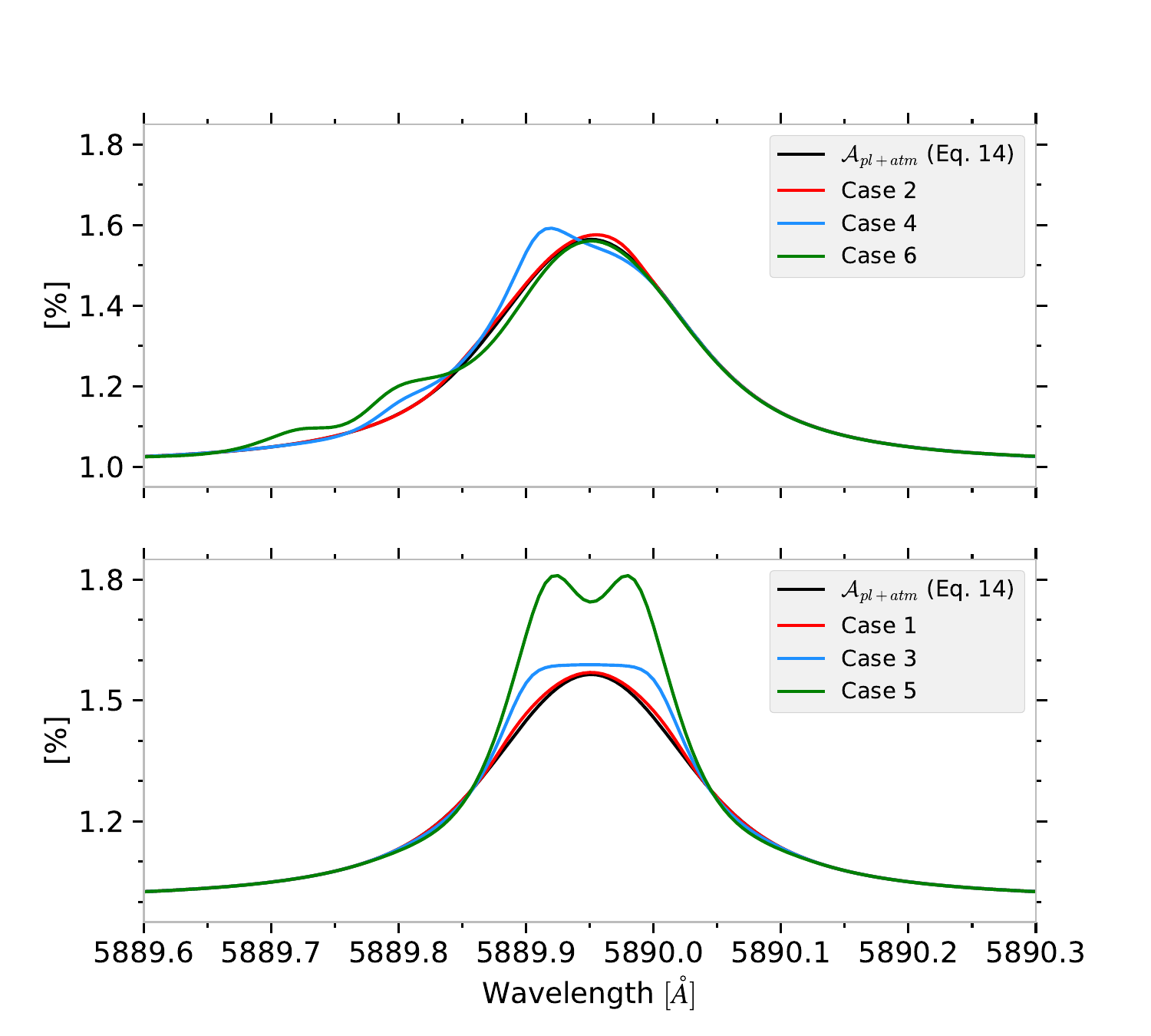}
    \caption{Comparison of the corrected absorption spectrum (Eq. \ref{eq:fout_fin2}) and the pure absorption signature (Eq. \ref{eq:fout_fin_simplified_b}) for an atmosphere with a radius of 1.5 times the planetary radius. Upper panel: When the planet is at $x,y = (-0.5;0)$. Lower panel: When the planet is at $x,y = (0;0)$, the centre of the stellar disc.}
    \label{fig:3}
\end{figure}

\begin{figure}
\hspace*{-5mm}
    \centering
    \includegraphics[width = 1.1\linewidth]{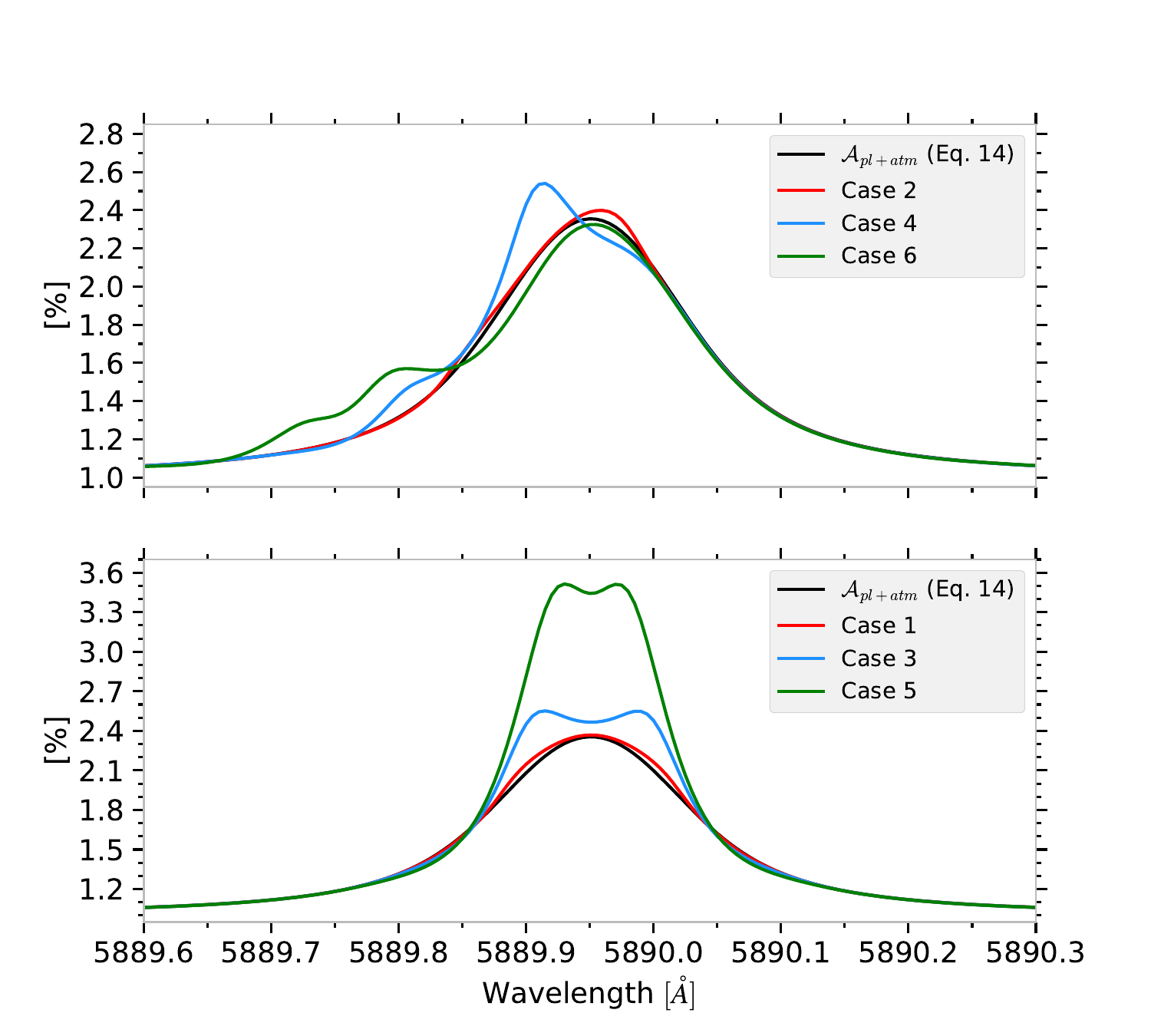}
    \caption{Comparison of the corrected absorption spectrum (Eq. \ref{eq:fout_fin2}) and the pure absorption signature (Eq. \ref{eq:fout_fin_simplified_b}) for an atmosphere with a radius of 2.0 times the planetary radius. Upper panel: When the planet is at $x,y = (-0.5;0)$. Lower panel: When the planet is at $x,y = (0;0)$, the centre of the stellar disc.}
    \label{fig:4}
\end{figure}

Figures \ref{fig:3} and \ref{fig:4} show the results of the simulations for the six different cases for an atmosphere of radius 1.5 times the planetary radius and 2.0 times the planetary radius, respectively. It is clear that when centre-to-limb variations and stellar rotation are included in the stellar spectrum, the result of Eq. \ref{eq:fout_fin2} diverges from the unaffected absorption spectrum of the atmosphere (Eq. \ref{eq:fout_fin_simplified_b}). With a line profile that varies, one can expect the derived values for the atmosphere's properties to be biased. Moreover, the line profile sometimes appears to have a shifted component (e.g. case 4 with a planet position not at the centre of the star), which could be wrongly interpreted as winds in the atmosphere when there are none. This means that correcting the absorption spectrum with the synthetic absorption spectrum of the planetary opaque body is not a reliable method.

The goal of dividing the absorption spectrum by the synthetic absorption spectrum of the planet is to correct for the POLDs and thus extract an unaffected absorption spectrum for the planetary atmosphere (Eq. \ref{eq:fout_fin_simplified_b}). Therefore, in order to estimate the bias left by the division on the true atmospheric signature, we used Eq. \ref{eq:fout_fin_simplified_b} to fit the result of Eq. \ref{eq:fout_fin2}. Here, only the temperature and the column density are free parameters. We used a simple least square algorithm for the fit. Table \ref{tab:3} shows the errors on the temperature and column density for the six cases for an atmosphere with a radius of 1.5 and 2.0 times the planetary radius. The values of the errors should only serve to demonstrate that resorting to division by the absorption spectrum of the planetary body to extract the atmospheric signal will lead to wrong estimates of the true properties of the planetary atmosphere. 

\begin{table}
\renewcommand{\arraystretch}{1.5}
\setlength{\tabcolsep}{12pt}
\caption{Errors on the parameters of the fit for the different simulations.}
\label{tab:3}
\centering
\begin{tabular}{ccc}
\hline
\hline
Case & Temperature [$\%$]& Column density [$\%$] \\
\hline
\multicolumn{3}{l}{Atmosphere of radius 1.5 times the planetary radius} \\
\hline
1 & 4 $(\searrow)$ & 2 $(\nearrow)$\\
2 & 4 $(\searrow)$ & 2 $(\nearrow)$\\
3 & 24 $(\searrow)$ & 11 $(\nearrow)$\\
4 & 11 $(\nearrow)$ & 5 $(\nearrow)$\\
5 & 87 $(\searrow)$ & 41 $(\nearrow)$\\
6 & 27 $(\nearrow)$ & 0.05 $(\nearrow)$\\
\\
\multicolumn{3}{l}{Atmosphere of radius 2.0 times the planetary radius} \\
\hline
1 & 7 $(\searrow)$& 3 $(\nearrow)$\\
2 & 8 $(\searrow)$& 2 $(\nearrow)$\\
3 & 40 $(\searrow)$& 19 $(\nearrow)$\\
4 & 20 $(\nearrow)$& 8 $(\nearrow)$\\
5 & ? $(\searrow)$& 78 $(\nearrow)$\\
6 & 66 $(\nearrow)$& 0.08 $(\nearrow)$\\
\hline
\end{tabular}
\begin{tablenotes}
\item[] \tiny{\textbf{Note.} $(\searrow)$ and $(\nearrow)$ indicate underestimates and overestimates, respectively. For case 5, with an atmospheric radius of 2 planetary radii, the temperature of the fit decreases to 1K. The temperature is thus not constrained in this case.}
\end{tablenotes}
\end{table}

\section{Conclusion}
We have demonstrated that correcting absorption spectra via division by a simulated absorption spectrum of a bare planet does not allow one to extract the pure atmospheric signal. 
Only in the most idealistic case (i.e. uniform stellar spectra and a constant atmospheric optical depth) can one hope to derive unbiased properties. In addition, we have shown how the atmospheric signal can be distorted when centre-to-limb variations in line profiles and stellar rotation are included in the stellar spectrum. We have shown that in the case of a typical Jupiter-sized planet with a typical sodium atmospheric signal, the retrieved properties of the atmosphere can significantly differ from the true value. Stellar rotation being one of the dominant effects that modify local spectra, the bias left after the correction is more significant with increasing $\rm v_{eq}\sin(i)$.\\ \indent These results are particularly crucial for studies of planetary atmospheric escape. Under hydrodynamic escape, an atmosphere can occupy a broad and sometimes asymmetric region (i.e. with escaping tails and bow shocks) around the planet \citep[e.g.][]{bourrier2015,bourrier2018,carolan2021,spake2021,tyler2024,gullysantiago2024}, leading the planet and the atmosphere to absorb very different spectra. One can thus expect the atmospheric signature to be distorted in the absorption spectrum if it is corrected with a simulated absorption spectrum of the bare planet. It is, however, worth noting that atmospheric escape studies are usually done on hydrogen and helium lines. The exact behaviour of such lines with respect to centre-to-limb variations might be different from that of the Na I doublet lines. Therefore, the impact of the problematic identified in this study might differ and would be worth investigating.\\
\indent On the contrary, the residual biases after the correction, although present, might become negligible when studying transit observations of Earth-sized planets. With a smaller size, the spectra absorbed by the opaque body and the atmosphere could be similar enough for the flux ratio (identified in Eq. \ref{eq:bias}) to be close to 1, potentially lessening the challenge of characterising these planets.\\
\indent This study was conducted with synthetic spectra, and the exact absorption spectrum of the planetary opaque body was used for the corrections of the absorption spectra. We have thus mitigated the impact of the correction on the residual atmospheric signal. However, with observations, the noise and the instrumental response can further affect the true atmospheric signal, and one might not simulate the exact absorption spectrum of the planet for the correction, which would lead to a possibly unforeseen bias. Moreover, all the simulations were performed with absolute flux spectra. This means there was no need for normalisation. As pointed out by \citet{dethier2023}, normalisation can affect the amplitude of the atmospheric absorption signature. Therefore, with non-flux-calibrated transit observations that, thus, require normalisation, the derived absorption signature after correction might be modified even more strongly than without normalisation, potentially leading to even more divergent results.\\ \indent Finally, in this work we decided to use absorption spectra to demonstrate our argument that a bias is left after the correction. However, this argument still holds when using transmission spectra. The equations given here would need to be adapted (see Appendix \ref{app:1}), and the exact residual bias would change but would still be present. A detailed description of how the bias behaves when correcting absorption and transmission spectra would require an in-depth study that explores the parameters affecting the spectra contained in Eq. \ref{eq:fout_fin2} (e.g. stellar rotation, planetary and atmosphere size, spectral type, specific lines, etc.), which is not the intent of this study.\\
\indent The method of correcting the high-resolution absorption spectrum of a transiting planet using a simulated absorption spectrum of the bare planet is, therefore, not a sustainable method for extracting unbiased properties of a transiting exoplanet atmosphere. We therefore advise using a different approach, one in which both a planet and an atmosphere are simulated during the transit to derive a self-consistent absorption spectrum \citep[e.g.][]{dethier2023}. In this way, the simulated absorption spectra can be directly compared to the absorption spectra reconstructed from observations without further corrections.
\begin{acknowledgements}
The author greatly thanks the referee for their relevant suggestions that allowed us to improve our study. Funded/Co-funded by the European Union (ERC, FIERCE, 101052347). Views and opinions expressed are however those of the author(s) only and do not necessarily reflect those of the European Union or the European Research Council. Neither the European Union nor the granting authority can be held responsible for them. This work was supported by FCT - Fundação para a Ciência e a Tecnologia through national funds and by these grants: UIDB/04434/2020; UIDP/04434/2020. \\
This work has made use of the VALD database, operated at Uppsala University, the Institute of Astronomy RAS in Moscow, and the University of Vienna.
\end{acknowledgements}

\bibliographystyle{aa}
\bibliography{bibliography}

\begin{thebibliography}{25}
\expandafter\ifx\csname natexlab\endcsname\relax\def\natexlab#1{#1}\fi

\bibitem[{{Alvarez} \& {Plez}(1998)}]{plez1998}
{Alvarez}, R. \& {Plez}, B. 1998, \aap, 330, 1109

\bibitem[{{Bourrier} {et~al.}(2015){Bourrier}, {Ehrenreich}, \& {Lecavelier des Etangs}}]{bourrier2015}
{Bourrier}, V., {Ehrenreich}, D., \& {Lecavelier des Etangs}, A. 2015, A\&A, 582, A65

\bibitem[{{Bourrier} {et~al.}(2018){Bourrier}, {Lecavelier des Etangs}, {Ehrenreich}, {Sanz-Forcada}, {Allart}, {Ballester}, {Buchhave}, {Cohen}, {Deming}, {Evans}, {Garc{\'\i}a Mu{\~n}oz}, {Henry}, {Kataria}, {Lavvas}, {Lewis}, {L{\'o}pez-Morales}, {Marley}, {Sing}, \& {Wakeford}}]{bourrier2018}
{Bourrier}, V., {Lecavelier des Etangs}, A., {Ehrenreich}, D., {et~al.} 2018, \aap, 620, A147

\bibitem[{{Carolan} {et~al.}(2021){Carolan}, {Vidotto}, {Hazra}, {Villarreal D'Angelo}, \& {Kubyshkina}}]{carolan2021}
{Carolan}, S., {Vidotto}, A.~A., {Hazra}, G., {Villarreal D'Angelo}, C., \& {Kubyshkina}, D. 2021, \mnras, 508, 6001

\bibitem[{{Casasayas-Barris} {et~al.}(2017){Casasayas-Barris}, {Palle}, {Nowak}, {Yan}, {Nortmann}, \& {Murgas}}]{Casasayasbarris2017}
{Casasayas-Barris}, N., {Palle}, E., {Nowak}, G., {et~al.} 2017, \aap, 608, A135

\bibitem[{{Casasayas-Barris} {et~al.}(2018){Casasayas-Barris}, {Pall\'e, E.}, {Yan, F.}, {Chen, G.}, {Albrecht, S.}, {Nortmann, L.}, {Van Eylen, V.}, {Snellen, I.}, {Talens, G. J. J.}, {Gonz\'alez Hern\'andez, J. I.}, {Rebolo, R.}, \& {Otten, G. P. P. L.}}]{casasayasbarris2018}
{Casasayas-Barris}, N., {Pall\'e, E.}, {Yan, F.}, {et~al.} 2018, A\&A, 616, A151

\bibitem[{{Casasayas-Barris} {et~al.}(2020){Casasayas-Barris}, {Pall\'e, E.}, {Yan, F.}, {Chen, G.}, {Luque, R.}, {Stangret, M.}, {Nagel, E.}, {Zechmeister, M.}, {Oshagh, M.}, {Sanz-Forcada, J.}, {Nortmann, L.}, {Alonso-Floriano, F. J.}, {Amado, P. J.}, {Caballero, J. A.}, {Czesla, S.}, {Khalafinejad, S.}, {L\'opez-Puertas, M.}, {L\'opez-Santiago, J.}, {Molaverdikhani, K.}, {Montes, D.}, {Quirrenbach, A.}, {Reiners, A.}, {Ribas, I.}, {S\'anchez-L\'opez, A.}, \& {Zapatero Osorio, M. R.}}]{casasayasbarris2020}
{Casasayas-Barris}, N., {Pall\'e, E.}, {Yan, F.}, {et~al.} 2020, A\&A, 635, A206

\bibitem[{Charbonneau {et~al.}(2000)Charbonneau, Brown, Latham, \& Mayor}]{charbonneau2000}
Charbonneau, D., Brown, T.~M., Latham, D.~W., \& Mayor, M. 2000, The Astrophysical Journal, 529, L45

\bibitem[{{Dethier} \& {Bourrier}(2023)}]{dethier2023}
{Dethier}, W. \& {Bourrier}, V. 2023, A\&A, 674, A86

\bibitem[{{Gully-Santiago} {et~al.}(2024){Gully-Santiago}, {Morley}, {Luna}, {MacLeod}, {Oklop{\v{c}}i{\'c}}, {Ganesh}, {Tran}, {Zhang}, {Bowler}, {Cochran}, {Krolikowski}, {Mahadevan}, {Ninan}, {Stef{\'a}nsson}, {Vanderburg}, {Zalesky}, \& {Zeimann}}]{gullysantiago2024}
{Gully-Santiago}, M., {Morley}, C.~V., {Luna}, J., {et~al.} 2024, \aj, 167, 142

\bibitem[{{Gustafsson} {et~al.}(2008){Gustafsson}, {Edvardsson, B.}, {Eriksson, K.}, {J\o{}rgensen, U. G.}, {Nordlund, \AA{}.}, \& {Plez, B.}}]{gustafsson2008}
{Gustafsson}, B., {Edvardsson, B.}, {Eriksson, K.}, {et~al.} 2008, A\&A, 486, 951

\bibitem[{{Heiter} {et~al.}(2021){Heiter}, {Lind}, {Bergemann}, {Asplund}, {Mikolaitis}, {Barklem}, {Masseron}, {de Laverny}, {Magrini}, {Edvardsson}, {J{\"o}nsson}, {Pickering}, {Ryde}, {Bayo Ar{\'a}n}, {Bensby}, {Casey}, {Feltzing}, {Jofr{\'e}}, {Korn}, {Pancino}, {Damiani}, {Lanzafame}, {Lardo}, {Monaco}, {Morbidelli}, {Smiljanic}, {Worley}, {Zaggia}, {Randich}, \& {Gilmore}}]{heiter2021}
{Heiter}, U., {Lind}, K., {Bergemann}, M., {et~al.} 2021, \aap, 645, A106

\bibitem[{Henry {et~al.}(2000)Henry, Marcy, Butler, \& Vogt}]{henry2000}
Henry, G.~W., Marcy, G.~W., Butler, R.~P., \& Vogt, S.~S. 2000, The Astrophysical Journal, 529, L41

\bibitem[{Keles {et~al.}(2024)Keles, Czesla, Poppenhaeger, Hauschildt, Carroll, Ilyin, Baratella, Steffen, Strassmeier, Bonomo, Gaudi, Henning, Johnson, Molaverdikhani, Nascimbeni, Patience, Reiners, Scandariato, Schlawin, Shkolnik, Sicilia, Sozzetti, Mallonn, Veillet, Wang, \& Yan}]{keles2024}
Keles, E., Czesla, S., Poppenhaeger, K., {et~al.} 2024, The PEPSI Exoplanet Transit Survey (PETS). V: New Na D transmission spectra indicate a quieter atmosphere on HD 189733b

\bibitem[{{Larsen} {et~al.}(2022){Larsen}, {Eitner}, {Magg}, {Bergemann}, {Moltzer}, {Brodie}, {Romanowsky}, \& {Strader}}]{larsen2022}
{Larsen}, S.~S., {Eitner}, P., {Magg}, E., {et~al.} 2022, \aap, 660, A88

\bibitem[{{Magg} {et~al.}(2022){Magg}, {Bergemann}, {Serenelli}, {Bautista}, {Plez}, {Heiter}, {Gerber}, {Ludwig}, {Basu}, {Ferguson}, {Gallego}, {Gamrath}, {Palmeri}, \& {Quinet}}]{magg2022}
{Magg}, E., {Bergemann}, M., {Serenelli}, A., {et~al.} 2022, \aap, 661, A140

\bibitem[{{Maguire} {et~al.}(2024){Maguire}, {Gibson}, {Nugroho}, {Fortune}, {Ramkumar}, {Gandhi}, \& {de Mooij}}]{maguire2024}
{Maguire}, C., {Gibson}, N.~P., {Nugroho}, S.~K., {et~al.} 2024, arXiv e-prints, arXiv:2404.10463

\bibitem[{{Maguire} {et~al.}(2023){Maguire}, {Gibson}, {Nugroho}, {Ramkumar}, {Fortune}, {Merritt}, \& {de Mooij}}]{maguire2023}
{Maguire}, C., {Gibson}, N.~P., {Nugroho}, S.~K., {et~al.} 2023, \mnras, 519, 1030

\bibitem[{{Masson} {et~al.}(2024){Masson}, {Vinatier}, {B{\'e}zard}, {L{\'o}pez-Puertas}, {Lamp{\'o}n}, {Debras}, {Carmona}, {Klein}, {Artigau}, {Dethier}, {Pelletier}, {Hood}, {Allart}, {Bourrier}, {Cadieux}, {Charnay}, {Cowan}, {Cook}, {Delfosse}, {Donati}, {Gu}, {H{\'e}brard}, {Martioli}, {Moutou}, {Venot}, \& {Wyttenbach}}]{masson2024}
{Masson}, A., {Vinatier}, S., {B{\'e}zard}, B., {et~al.} 2024, arXiv e-prints, arXiv:2406.09225

\bibitem[{{Nugroho} {et~al.}(2020){Nugroho}, {Gibson}, {de Mooij}, {Watson}, {Kawahara}, \& {Merritt}}]{nugroho2020}
{Nugroho}, S.~K., {Gibson}, N.~P., {de Mooij}, E. J.~W., {et~al.} 2020, \mnras, 496, 504

\bibitem[{{Plez}(2012)}]{plez2012}
{Plez}, B. 2012, {Turbospectrum: Code for spectral synthesis}

\bibitem[{Ryabchikova {et~al.}(2015)Ryabchikova, Piskunov, Kurucz, Stempels, Heiter, Pakhomov, \& Barklem}]{Ryabchikova2015}
Ryabchikova, T., Piskunov, N., Kurucz, R.~L., {et~al.} 2015, Physica Scripta, 90, 054005

\bibitem[{{Spake} {et~al.}(2021){Spake}, {Oklop{\v{c}}i{\'c}}, \& {Hillenbrand}}]{spake2021}
{Spake}, J.~J., {Oklop{\v{c}}i{\'c}}, A., \& {Hillenbrand}, L.~A. 2021, \aj, 162, 284

\bibitem[{{Tyler} {et~al.}(2024){Tyler}, {Petigura}, {Oklop{\v{c}}i{\'c}}, \& {David}}]{tyler2024}
{Tyler}, D., {Petigura}, E.~A., {Oklop{\v{c}}i{\'c}}, A., \& {David}, T.~J. 2024, \apj, 960, 123

\bibitem[{{Yan} \& {Henning}(2018)}]{yan2018}
{Yan}, F. \& {Henning}, T. 2018, Nature Astronomy, 2, 714

\end{thebibliography}
 
\begin{appendix}
\section{Equations of Sect \ref{sec:3} for transmission spectra}
\label{app:1}
We remind the reader that, even though a transmission spectrum is defined as $\mathcal{T} = 1 - \mathcal{A}$, the corrected transmission spectrum is not equal to 1 minus the corrected absorption spectrum. Rather, it is equal to $\rm \mathcal{T}_{corr} = \frac{\mathcal{T}_{pl+atm}}{\mathcal{T}_{pl}} = \frac{1 - \mathcal{A}_{pl+atm}}{1 - \mathcal{A}_{pl}}$. 

As such, the equations provided in the main document for the absorption spectra cannot simply be transformed using 1 minus a given equation for the absorption spectrum. Below, we give the equivalent of Eqs. \ref{eq:fout_fin} to \ref{eq:fout_fin_simplified_b} from Sect. \ref{sec:3} in the case of transmission spectra instead of absorption spectra.\\
\indent The transmission spectrum of the planet and its atmosphere is written as
\begin{equation}
\begin{split}
& \rm \mathcal{T}_{pl+atm} = \frac{\overbrace{\displaystyle\sum\limits_j^J \text{F}_j}^\text{{\large$\text{F}_\text{unocc}$}}   \ +\overbrace{\displaystyle\sum\limits_l^L \text{F}_l\ e^{-\tau_l}}^\text{{\large $\text{F}_\text{trans atm}$}}}{\displaystyle\sum_{i=1}^n \text{F}_i},
\end{split}
\label{eq:fout_fin_T}
\end{equation}
where the $j$ index runs on the cells that are not occulted, neither by the planet nor the atmosphere. $\text{F}_\text{trans atm}$ designates the flux transmitted through the atmosphere.\\ \indent The transmission spectrum of the planet is written as\begin{equation}
\begin{split}
& \rm \mathcal{T}_{pl} = \frac{\overbrace{\displaystyle\sum\limits_j^J \text{F}_j}^\text{{\large$\text{F}_\text{unocc}$}} \ +\overbrace{\displaystyle\sum\limits_l^L \text{F}_l}^\text{{\large $\text{F}_\text{unocc atm}$}}}{\displaystyle\sum_{i=1}^n \text{F}_i},
\end{split}
\label{eq:fout_fin1_T}
\end{equation}
where $\text{F}_\text{unocc atm}$ designates the flux behind the atmospheric annulus.
The corrected transmission spectrum is written as
\begin{equation}
\begin{split}
 \rm \dfrac{\mathcal{T}_{pl+atm}}{\mathcal{T}_{pl}} =& \frac{\displaystyle\sum\limits_j^J \text{F}_j   \ +\displaystyle\sum\limits_l^L \text{F}_l\ e^{-\tau_l}}{\displaystyle\sum\limits_j^J \text{F}_j \ +\displaystyle\sum\limits_l^L \text{F}_l}.
\end{split}
\label{eq:fout_fin2_T}
\end{equation}
Under the assumptions of uniform stellar spectra and a constant atmospheric optical depth, we simplify Eq. \ref{eq:fout_fin2_T} as
\begin{equation}
\begin{split}
 \rm \dfrac{\mathcal{T}_{pl+atm}}{\mathcal{T}_{pl}} =& \frac{\rm S_* - S_{pl} - S_{atm}    \ +S_{atm}\ e^{-\tau}}{\rm S_* - S_{pl}}\\
=&  1 + (e^{-\tau} - 1) \frac{\rm S_{atm}}{\rm S_* - S_{pl}}.
\end{split}
\label{eq:fout_fin4_T}
\end{equation}
Under the same assumptions, Eq. \ref{eq:fout_fin_T} is written as\begin{equation}
\begin{split}
& \rm \mathcal{T}_{pl+atm} = \frac{\rm S_{*} - S_{pl}  \ + S_{atm}\ (e^{-\tau}-1)}{\rm S_*},
\end{split}
\label{eq:fout_fin_simplified_T}
\end{equation}
which becomes equivalent to Eq. \ref{eq:fout_fin4_T} after the following operation,
\begin{equation}
\begin{split}
& \rm \mathcal{T}_{pl+atm} \times  \frac{\rm S_{*}}{\rm S_* - S_{pl}} = 1+\ (e^{-\tau} - 1)\frac{S_{atm}}{\rm S_* - S_{pl}}.
\end{split}
\label{eq:fout_fin_simplified_b_T}
\end{equation}

\noindent Finally, the bias is defined as Eq. \ref{eq:fout_fin2_T} minus Eq. \ref{eq:fout_fin_simplified_b_T},
\begin{equation}
\begin{split}
 \rm \mathcal{B_T} & \equiv  \rm\dfrac{\mathcal{T}_{pl+atm}}{\mathcal{T}_{pl}} -\left.\dfrac{\mathcal{T}_{pl+atm}}{\mathcal{T}_{pl}}\right|_{sim}  \\ 
 & = \frac{\displaystyle\sum\limits_j^J \text{F}_j   \ +\displaystyle\sum\limits_l^L \text{F}_l\ e^{-\tau_l}}{\displaystyle\sum\limits_j^J \text{F}_j \ +\displaystyle\sum\limits_l^L \text{F}_l} - 1 - \ \frac{\displaystyle\sum\limits_l^L (e^{-\tau_l} - 1) \rm \ S_l}{\rm S_* - S_{pl}} \\
 & = \frac{ \rm F_* - F_{pl} - F_{atm} \ +\displaystyle\sum\limits_l^L \text{F}_l\ e^{-\tau_l}}{ \rm F_* - F_{pl}} - 1 - \ \frac{\displaystyle\sum\limits_l^L (e^{-\tau_l} - 1) \rm \ S_l}{\rm S_* - S_{pl}} \\
 & = \frac{ \rm \displaystyle\sum\limits_l^L \text{F}_l\ e^{-\tau_l} - F_{atm}}{ \rm F_* - F_{pl}} - \ \frac{\displaystyle\sum\limits_l^L (e^{-\tau_l} - 1) \rm \ S_l}{\rm S_* - S_{pl}} \\
  & = \frac{ \rm \displaystyle\sum\limits_l^L \text{F}_l\ (e^{-\tau_l} - 1)}{ \rm F_* - F_{pl}} - \ \frac{\displaystyle\sum\limits_l^L (e^{-\tau_l} - 1) \rm \ S_l}{\rm S_* - S_{pl}} \\
  & = \rm \displaystyle\sum\limits_l^L\ (e^{-\tau_l} - 1)\left[\frac{\text{F}_l} { \rm F_* - F_{pl}} - \ \frac{ \rm \ S_l}{\rm S_* - S_{pl}}\right]
\end{split}
\label{eq:biast}
.\end{equation}
\begin{figure}
\hspace*{-3mm}
    \centering
    \includegraphics[width = 1.1\linewidth]{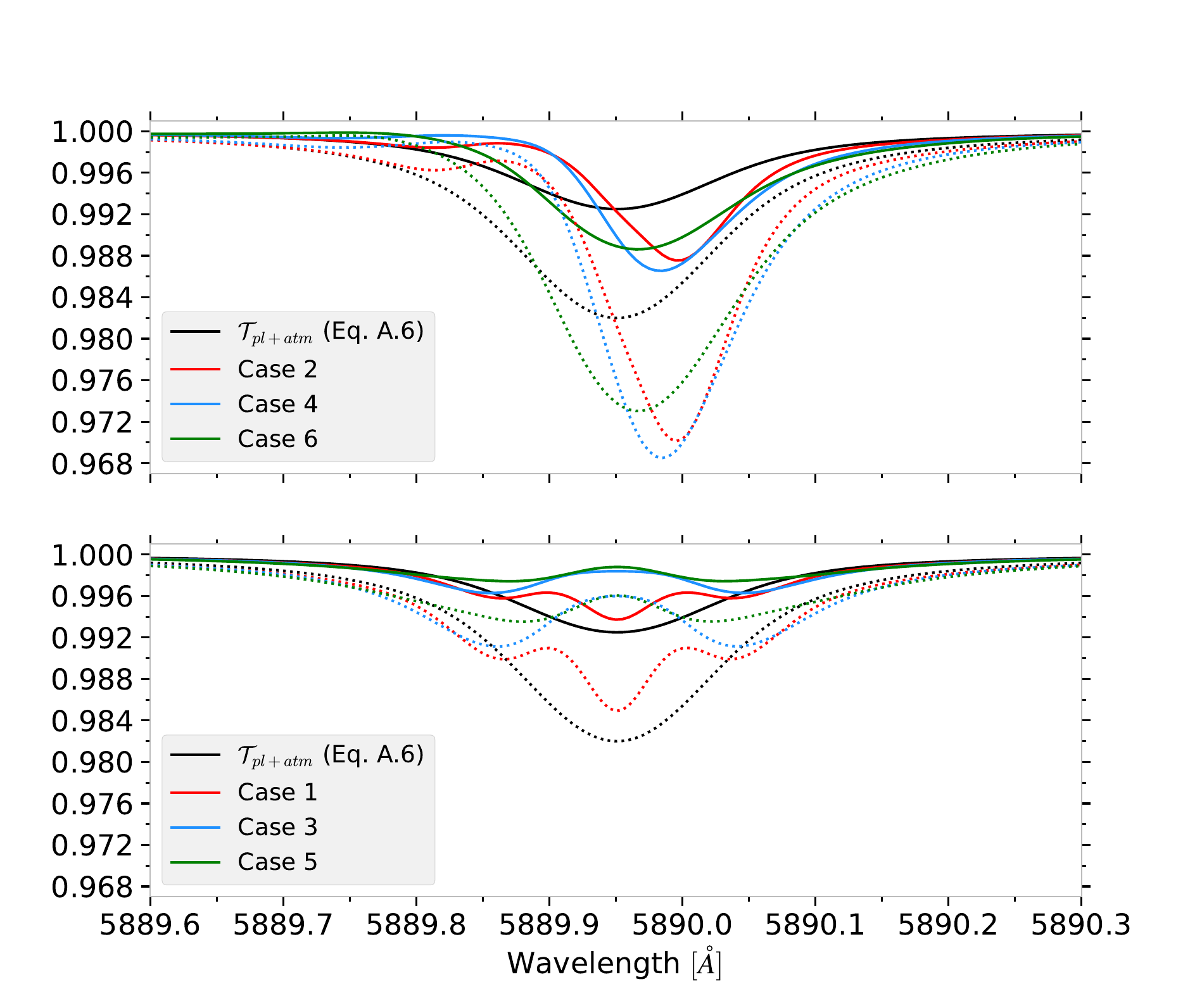}
    \caption{Comparison of the corrected transmission spectrum (Eq. \ref{eq:fout_fin2_T}) and the pure transmission signature (Eq. \ref{eq:fout_fin_simplified_b_T}) for an atmosphere with a radius of 1.5 times the planetary radius (plain curves) and 2.0 times the planetary radius (dotted curves). Upper panel: When the planet is at $x,y = (-0.5;0)$. Lower panel: When the planet is at $x,y = (0;0)$, the centre of the stellar disc.}
    \label{fig:A}
\end{figure}
\end{appendix}
\end{document}